\documentclass[aps,pre,showpacs,preprintnumbers,superscriptaddress]{revtex4}
\usepackage{epsfig,graphics,amssymb,amsmath,subeqnarray}
\usepackage{color}

\def\p{\partial}\def\u{{\bf u}}\def\e{{\bf e}}\def\btau{\boldsymbol{\tau}}\def\bgam{\boldsymbol{\gamma}}\def\gamd{\dot\bgam}\def\De{{\rm De}}
\def\n{{\bf n}}\def\t{{\bf t}}\def\dgamd{\stackrel{\triangledown}{\gamd}}
\def\bs{{\boldsymbol{\sigma}}}\def\dbtau{\stackrel{\triangledown}{\btau}}
\def\Real{{\rm Re}}

\def\varepsilon{\epsilon}

\begin{document}

\title{Pumping by flapping in a viscoelastic fluid}
\author{On Shun Pak}
\affiliation{Department of Mechanical and Aerospace Engineering, 
University of California San Diego, 9500 Gilman Dr., La Jolla CA 92093-0411, USA.}
\author{Thibaud Normand}
\affiliation{D\'epartment de M\'ecanique, Ecole Polytechnique, 91128 Palaiseau Cedex, France.}
\author{Eric Lauga}
\email{elauga@ucsd.edu}
\affiliation{Department of Mechanical and Aerospace Engineering, 
University of California San Diego, 9500 Gilman Dr., La Jolla CA 92093-0411, USA.}
\date{\today}

\begin{abstract}
In a world without inertia, Purcell's scallop theorem states that in a Newtonian fluid a time-reversible motion cannot produce any net force or net flow. Here we consider the extent to which  the nonlinear rheological behavior of viscoelastic fluids can be exploited to break the constraints of the scallop theorem in the context of fluid pumping. By building on previous work focusing on force generation, we consider a simple,  biologically-inspired geometrical example of a flapper in a polymeric (Oldroyd-B) fluid, and calculate asymptotically the time-average net fluid flow produced by the reciprocal flapping motion. The net flow occurs at fourth order in the flapping amplitude, and suggests the possibility of transporting polymeric fluids using reciprocal motion in simple geometries even in the absence of inertia. The induced flow field and pumping performance are characterized and optimized analytically. Our results may be useful in the design of micro-pumps handling complex fluids.  
 \end{abstract}


\maketitle

%%%%%%%%%%%%%%%%%%%%
%%%%%% GENERAL
%%%%%%%%%%%%%%%%%%%%
\section{\label{sec:introduction}Introduction}
Imagining oneself attempting to swim in a pool of viscous honey, it is not hard to anticipate that, because of the high fluid viscosity, our usual swimming strategy consisting of imparting momentum to the surrounding fluid will not be effective.  The world microorganisms inhabit is physically analogous to this situation \cite{purcell77}. As a result, microorganisms have evolved strategies which exploit the only physical force available to them -- namely fluid drag -- to propel themselves or generate net fluid transport. The success of these propulsion strategies is vital in many biological processes, including bacterial infection, spermatozoa locomotion and reproduction, and ciliary transport \cite{braybook}.  Advances in micro- and nano- manufacturing technology have also allowed scientists to take inspiration from these locomotion strategies and design micropumps \cite{leoni08} and microswimmers \cite{dreyfus}. 
 
The fundamental physics of small-scale locomotion in simple (Newtonian) fluids is well understood \cite{brennen,childress,lauga2}.  In contrast, and although most biological fluids are non-Newtonian, many basic questions  remain unanswered regarding the mechanics of  motility in complex fluids. Since they usually include biopolymers, most biological fluids of interest display rheological properties common to both fluids (they flow and dissipate energy) and solids (they can store energy). Examples include the airway mucus, which acts as a renewable and transportable barrier along the airways of the lungs to guard against inhaled particulates and toxic substances \cite{samet}, as well as cervical mucus, which is important for the survival and transport of sperm cells \cite{yudin}. The influence of viscoelasticity of the fluid on cell locomotion has been experimentally quantified by a number of studies \cite{shukla, katz1, katz2, katz3, rikmenspoel, ishijima, suarez}, including  the change in the  waveform, structure, and swimming path of spermatozoa in both synthetic polymer solutions and biological mucus \cite{fauci3}. Gastropod mucus is another common non-Newtonian biofluid, which is useful for adhesive locomotion, and its physical and rheological properties have been measured \cite{denny,ewoldt07,ewoldt08}. Modeling-wise, different constitutive models have been employed to study locomotion in complex fluids (see the short review in Ref.~\cite{lauga1}). Among these models, the Oldroyd-B constitutive equation is the most popular, both because of its simplicity and the fact that it can be derived exactly from  kinetic theory by modeling the fluid as a dilute solutions of elastic (polymeric) dumbbells \cite{oldroyd, bird1, bird2, johnson}.  Recent quantitative studies have suggested that microorganisms swimming by propagating waves along their flagella have a smaller propulsion speed in a polymeric fluid than in a Newtonian fluid \cite{lauga1, fu}. Likewise, a smaller net flow is generated by the ciliary transport of a polymeric fluid than a Newtonian fluid. Specifically, Lauga \cite{lauga1} considered the problem with a prescribed beating pattern along the flagellum, while Fu and Powers \cite{fu} prescribed the internal force distribution instead; both studies suggest that viscoelasticity tends to decrease the propulsion speed.

In a Newtonian fluid, Purcell's scallop theorem states that swimming and pumping in the absence of inertia can only be achieved by  motions or body deformations which are not identical under a time-reversal symmetry (so-called ``non-reciprocal'' motion)  \cite{purcell77}. This poses of course an interesting challenge in designing artificial swimmers and pumps in simple fluids, which has recently been addressed  theoretically and experimentally (see the review in Ref.~\cite{lauga2}). The question we are addressing in this paper is the extent to which the  scallop theorem holds in complex fluids. Because polymeric fluids display nonlinear rheological properties such as shear-dependance or normal-stress differences \cite{bird1, bird2}, in general reciprocal motions are effective in polymeric fluids \cite{lauga3}. New propulsion and transport methods can therefore be designed on small scales to specifically  take advantage of  the intrinsic  nonlinearities  of the fluid. The goal of this paper is to study such a system in the context of fluid pumping with  a simplified geometrical setup where the pumping performance can be characterized analytically.

For simple flow geometries, it is not obvious a priori whether a simple oscillatory forcing of a nonlinear fluid leads to a net (rectified) flow. For example, for all Oldroyd-like fluids, a sinusoidally-forced Couette flow leads to zero time-averaged flow \cite{bird1}. In previous work  \cite{norman}, we considered a biologically-inspired geometric example of a semi-infinite flapper performing reciprocal (sinusoidal) motion in a viscoelastic (Oldroyd-B) fluid in the absence of inertia. We showed explicitly that the reciprocal motion generates a net force on the flapper occurring at second order in the flapping amplitude, and disappearing in the Newtonian limit as dictated by the scallop theorem. 
However, there was no time-average flow accompanying the net force generation at second order \cite{norman}. Here, we report on the discovery of a net fluid flow produced by the reciprocal flapping motion in an Oldroyd-B fluid. The net flow transport is seen to occur at fourth order in the flapping amplitude, and is due to normal-stress differences. The dependence of the pumping performance  on the actuation and material parameters is characterized analytically, and the optimal pumping rate is determined numerically.  
Through this example, we therefore demonstrate explicitly the breakdown of the scallop theorem in complex fluids in the context of fluid pumping, and suggest the possibility of exploiting intrinsic viscoelastic  properties of the medium for fluid  transport  on small scales.

The geometric setup in this paper is motivated by the motion of cilia-like biological appendages. Cilia are short flagella beating collaboratively to produce locomotion or fluid transport \cite{gibbons,brennen}. For example, cilia cover the outer surface of microorganisms such as \textit{Paramecium} for self-propulsion. They are also present  along our respiratory tracts to sweep up dirt and mucus and along the oviduct to transport the ova. Our setup is also relevant to the rigid projections attached to the flagellum of \textit{Ochromonas}, known as mastigonemes, which protrude from the flagellum into the fluid \cite{brennen}. As waves propagate along the flagellum, the mastigonemes are flapped back-and-forth passively through the fluid, a process which can lead to a change in the direction of propulsion of the organism \cite{jahn, holwill, brennen2}. 

Our study is related to the phenomenon known as steady (or ``acoustic'') streaming in the inertial realm \cite{faraday, rayleigh, schlichting, riley, chang, james, rosenblat, bohme, bagchi, frater, frater2, goldstein, chang2, chang3}, which has a history of almost two centuries after being first discovered by Faraday \cite{faraday}. Under oscillatory boundary conditions, as in the presence of an acoustic wave or the periodic actuation of a solid body in a fluid, migration of fluid particles occur in an apparently purely oscillating flow, manifesting the presence of nonlinear inertial terms in the equation of motion. This phenomenon occurs in both Newtonian and non-Newtonian fluids \cite{chang, james, rosenblat, bohme, bagchi, frater, frater2, goldstein, chang2, chang3}. In particular, it was found that the elasticity of a polymeric fluid can lead to a reversal of the net flow direction  \cite{chang, james, rosenblat, bohme}. As expected from the scallop theorem, no steady streaming phenomenon can occur in a Newtonian fluid in the absence of inertia. However, as will be shown in this paper, the nonlinear rheological properties of viscoelastic fluids alone can lead to steady streaming. In other words, we consider here a steady streaming motion arising purely from the viscoelastic effects of the fluid, ignoring any influence of inertia.

Recently, polymeric solutions have been shown to be useful in constructing microfluidic devices such as flux stabilizers, flip-flops and rectifiers \cite{groisman1, groisman2}. By exploiting the nonlinear rheological properties of the fluid and geometrical asymmetries in the micro-channel, microfluidic memory and control have been demonstrated without the use of external electronics and interfaces, opening the possibility of more complex integrated microfluidic circuit and other medical applications \cite{groisman1}. In the setup we study here, we do not introduce any geometrical asymmetries  and exploit solely the non-Newtonian rheological properties of the polymeric fluid for microscopic fluid transport.

The structure of the paper is the following. In \S\ref{sec:formulation}, the flapping problem is formulated with the geometrical setup, governing equations, nondimensionalization and the boundary conditions. In \S\ref{sec:analysis}, we present the asymptotic calculations up to the fourth order (in flapping amplitude), where the time-average flow is obtained. We then characterize analytically the net flow in terms of the streamline pattern, directionality and vorticity distribution (\S\ref{sec:results}). Next, we study the optimization of the flow with respect to the Deborah number (\S\ref{sec:optimization}). Our results are finally discussed in \S\ref{sec:discussion}.

\section{\label{sec:formulation}Formulation}
\subsection{Geometrical setup}
\begin{figure}
\begin{center}
\includegraphics[width=0.4\textwidth]{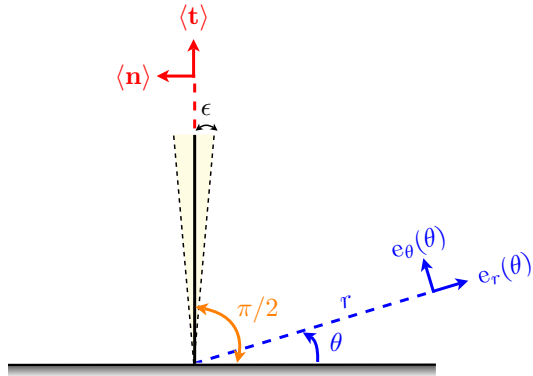}
\end{center}
\caption{\label{fig:setup}Geometrical setup and notations for the flapping calculation. A semi-infinite plane flaps sinusoidally with small amplitude $\epsilon$ around an average position at right angle with an infinite wall.}
\end{figure}

In this paper, we consider a semi-infinite two-dimensional plane flapping sinusoidally with small amplitude in a viscoelastic fluid. The average position of the flapper is situated perpendicularly to a flat wall  with its hinge point fixed in space (see Fig.~\ref{fig:setup}). The flapper is therefore able to perform reciprocal motion with only one degree of freedom by flapping back-and-forth. Such a setup is directly relevant to the unsteady motion of cilia-like biological appendages (see \S\ref{sec:introduction}). 

It is convenient to adopt planar polar coordinates system for this geometrical setup. The instantaneous position of the flapper is described by the azimuthal angle $\theta(t) = \pi/2 + \epsilon \Theta(t)$, where $\Theta(t)$ is an order one oscillatory function of time and $\epsilon$ is a parameter characterizing the amplitude of the flapping motion. The polar vectors $\bf{e}_r(\theta)$ and $\bf{e}_\theta(\theta)$ are functions of the azimuthal angle, and the velocity field $\u$ is expressed as $\u = u_r \e_r + u_\theta \e_\theta$.
In this work, we derive the velocity field in the the domain ($0 \leq \theta \leq \pi/2$) in the asymptotic limit of small flapping amplitude, {\it i.e.} $\epsilon \ll 1$; the time-averaged flow in the domain ($\pi/2 \leq \theta \leq \pi$) can then be deduced by symmetry.

\subsection{Governing equations}
We assume the flow to be incompressible and the Reynolds number of the fluid motion to be small, {\it i.e.} we neglect any inertial effects. Denoting  the pressure field as $p$ and the deviatoric stress tensor as  $\btau$, the continuity equation and Cauchy's equation of motion are respectively
\begin{align}
\boldsymbol{\boldsymbol{\nabla}} \cdot \u &=0,\\
\boldsymbol{\boldsymbol{\nabla}}p&=\boldsymbol{\boldsymbol{\nabla}} \cdot \btau.
\end{align}

We also require constitutive equations, which relate stresses and kinematics of the flow, to close the system of equations. For polymeric fluids, the deviatoric stress may be decomposed into two components, $\btau = \btau^s+\btau^p$, where $\btau^s$ is the Newtonian contribution from the solvent and $\btau^p$ is the polymeric stress contribution. For the Newtonian contribution, the constitutive equation is simply given by $\btau^s = \eta_s \gamd$, where $\eta_s$ is the solvent contribution to the viscosity and $\gamd = \boldsymbol{\boldsymbol{\nabla}} \u + {}^t \boldsymbol{\boldsymbol{\nabla}}\u$. For the polymeric contribution, many models have been proposed to relate the polymeric stress to kinematics of the flow \cite{oldroyd, bird1, bird2, johnson}. 
We consider here the classical Oldroyd-B model,  where the polymeric stress, $\btau^p$, satisfies the upper-convected Maxwell equation
\begin{align}\label{eq:upperMaxwell}
\btau^p + \lambda \stackrel{\triangledown}{\btau^p} = \eta_p \gamd, 
\end{align}
where $\eta_p$ is the polymer contribution to the viscosity and $\lambda$ is the polymeric relaxation time. The upper-convected derivative for a tensor $\bf{A}$ is defined as 
\begin{align}
\stackrel{\triangledown}{\bf{A}} = \frac{\p \bf{A}}{\p t} +\u \cdot  \boldsymbol{\boldsymbol{\nabla}} \bf{A} -\left({}^t \boldsymbol{\boldsymbol{\nabla}} \u \cdot \bf{A}+\bf{A} \cdot \boldsymbol{\boldsymbol{\nabla}} \u  \right),
\end{align}
and represents the rate of change of $\bf{A}$ in the frame of translating and deforming with the fluid. From Eq.~(\ref{eq:upperMaxwell}), we can obtain the Oldroyd-B constitutive equation for the total stress, $\btau$, as given by
\begin{align}
\btau + \lambda_1 \dbtau = \eta \left( \gamd + \lambda_2 \dgamd \right),
\end{align}
where $\eta=\eta_s+\eta_p$, $\lambda_1=\lambda$, and $\lambda_2 = \eta_s \lambda / \eta$. Here, $\lambda_1$ and $\lambda_2$ are the relaxation and retardation times of the fluid respectively. The relaxation time is the typical decay rate of stress when the fluid is at rest, and the retardation time measures the decay rate of residual rate of strain when the fluid is stress-free \cite{bird1, bird2}. It can be noted that $\lambda_2 < \lambda_1$, and both are zero for a Newtonian fluid.

\subsection{Nondimensionalization}
Periodic flapping motion with angular frequency $\omega$ is considered in this paper. Therefore, we nondimensionalize shear rates by $\omega$ and stresses by $\eta \omega$. Lengths are nondimensionalized by some arbitrary length scale along the flapper. Under these nondimensionalizations, the dimensionless equations are given by
\begin{subequations}\label{eq:govern}
\begin{align}
\boldsymbol{\boldsymbol{\nabla}} \cdot \u &=0,\\
\boldsymbol{\boldsymbol{\nabla}}p&=\boldsymbol{\boldsymbol{\nabla}} \cdot \btau, \label{eq:mecheq}\\  
\btau + \De_1 \dbtau &= \gamd + \De_2 \dgamd, 
\end{align} 
\end{subequations}
where $\De_1=\lambda_1 \omega$ and $\De_2 = \lambda_2 \omega$ are defined as the two Deborah numbers and we have adopted the same symbols for convenience.

\subsection{Boundary conditions}
The boundary condition in this problem can be simply stated; on the flat wall ($\theta = 0$), we have the no-slip and the no-penetration  boundary conditions. In vector notation, we have therefore
\begin{align}\label{eq:bc1}
\u (r, \theta = 0) = \bf{0}
\end{align}
along the flat wall.

Along the flapper, we also have the no-slip condition, $u_r (r, \theta=\pi/2+\epsilon \Theta(t))=0$. The other boundary condition imposed on the fluid along the flapper is given by the rotation of the flapper, $u_\theta (r, \theta=\pi/2+\epsilon \Theta(t))=r \Omega(t)$, where $\Omega(t)=\epsilon \dot{\Theta}$. In vector notation, we have then
\begin{align}\label{eq:bc2}
\u(r, \theta = \pi/2+\epsilon \Theta (t)) = r \Omega(t) \e_\theta.
\end{align}

\section{\label{sec:analysis}Analysis}
Noting that a two-dimensional setup is considered, the continuity equation, $\boldsymbol{\boldsymbol{\nabla}} \cdot \u =0$, is satisfied by introducing the streamfunction $\Psi (r, \theta)$ such that $u_r=\left(\p \Psi / \p \theta \right)/r$ and $u_\theta = - \p \Psi / \p r$. The instantaneous position of the flapper is described by the function $\theta = \pi/2+\epsilon \Theta(t)$, and we consider here a simple reciprocal flapping motion with $\Theta(t) = \cos t$. Since small amplitude flapping motion ($\epsilon \ll 1$) is considered, we will perform the calculations perturbatively in the flapping amplitude and seek perturbation expansions of the form
\begin{align}\label{eq:perturb}
\{\u, \Psi, \btau, p, \bs \} & =\  \epsilon \{\u_1, \Psi_1, \btau_1, p_1, \bs_1 \} + \epsilon^2 \{\u_2, \Psi_2, \btau_2, p_2, \bs_2 \} + \ldots,
\end{align}
where $\bs = -p \bf{1} +\btau$ is the total stress tensor and all the variables in  Eq.~(\ref{eq:perturb}) are defined in the time-averaged domain $0 \leq \theta \leq \pi/2$. Since a domain-perturbation expansion is performed, careful attention has to be paid on the distinction between instantaneous and average geometry. Recall that the polar vectors $\e_r(\theta(t))$ and $\e_\theta (\theta(t))$ are functions of the azimuthal angle which oscillates in time. To distinguish the average geometry, we denote $\langle \textbf{t} \rangle = \e_r (\pi/2)$ and $\langle \textbf{n} \rangle = \e_\theta (\pi/2)$ as the average directions along and perpendicular to the flapper axis (See Fig.~\ref{fig:setup}). In this paper, $\langle  \dots \rangle$ denotes time-averaging.

In addition, we employ Fourier notation to facilitate the calculations. In Fourier notation, the actuation becomes $\Theta = \Real \{e^{i t} \}$ and $\dot{\Theta} = \Real \{i e^{i t} \}$. Because of the quadratic nonlinearities arising from boundary conditions and the constitutive modeling, the velocity field can be Fourier decomposed into the anticipated form
\begin{subequations}
\begin{align}
\u_1 &= \Real \{\tilde{\u}_1 e^{it} \},\label{eq:fourieru1}\\
\u_2 &= \Real \{\tilde{\u}^{(0)}_2+\tilde{\u}^{(2)}_2 e^{2it} \}, \\
\u_3 &= \Real \{\tilde{\u}^{(1)}_3 e^{it}+\tilde{\u}^{(3)}_3 e^{3it} \},\\
\u_4 &= \Real \{\tilde{\u}^{(0)}_4+\tilde{\u}^{(2)}_4 e^{2it}+\tilde{\u}^{(4)}_4 e^{4it} \},
\end{align}
\end{subequations}
with similar decomposition and notation for all other vector and scalar fields.

We now proceed to analyze Eq.~(\ref{eq:govern}) order by order, up to the fourth order, where the time-average fluid flow occurs. The boundary conditions, Eqs.~(\ref{eq:bc1}) and (\ref{eq:bc2}), are also expanded order by order about the average flapper position using Taylor expansions.

\subsection{First-order solution}
\subsubsection{Governing equation}
The first-order Oldroyd-B relation is given by 
\begin{align}\label{eq:constitutive1}
\btau_1+\De_1 \frac{\p \btau_1}{\p t} = \gamd_1+\De_2 \frac{\p \gamd_1}{\p t},
\end{align}
which in Fourier space becomes
\begin{align}
\tilde\btau_1 = \frac{1+i\De_2}{1+i\De_1}\tilde\gamd_1. \label{eq:constitutiveF1}
\end{align}
We then note that we have $\boldsymbol{\nabla} \times \boldsymbol{\nabla} \cdot \btau =0$ by taking the curl of Eq.~(\ref{eq:mecheq}). Therefore, we take the divergence and then curl of Eq.~(\ref{eq:constitutiveF1}) to eliminate the stress and obtain the equation for the first-order streamfunction
\begin{align}
\boldsymbol{\nabla}^4 \tilde\Psi_1&=0.
\end{align}

\subsubsection{Boundary conditions}
At $\theta = \pi/2$, the boundary condition at this order is given by
\begin{align}
\u_1 = r \dot{\Theta} \langle \bf{n} \rangle \label{eq:order1bc},
\end{align}
which becomes
\begin{align}
\tilde\u_1 &= i r \langle \bf{n} \rangle,
\end{align}
upon Fourier transformation. We also have the no-slip and no-penetration boundary condition at $\theta=0$.

\subsubsection{Solution}
The solution satisfying the above equation and boundary conditions is given by
\begin{subequations}
\begin{align}
\tilde\Psi_1&=\frac{i r^2}{4}\left( \cos2\theta -1\right), \\
\tilde{{u}}{}_{1r}&=-\frac{i r}{2}\sin2\theta, \\
\tilde{{u}}_{1\theta}&=\frac{i r}{2}\left( 1-\cos2\theta \right).
\end{align}
\end{subequations}

\subsection{Second-order solution}
\subsubsection{Governing equation}
The second-order Oldroyd-B relation is given by 
\begin{align}\label{eq:constitutive2}
&\left( 1+\De_1 \frac{\p }{\p t}\right) \btau_2 -\left( 1+\De_2 \frac{\p }{\p t}\right) \gamd_2 \notag \\
&= \  \De_2 \left[\u_1 \cdot \boldsymbol{\nabla} \gamd_1- \left(^{t} \boldsymbol{\nabla} \u_1 \cdot \gamd_1+\gamd_1 \cdot \boldsymbol{\nabla} \u_1 \right) \right] \notag \\
& - \De_1  \left[\u_1 \cdot \boldsymbol{\nabla} \btau_1- \left(^{t} \boldsymbol{\nabla} \u_1 \cdot \btau_1+\btau_1 \cdot \boldsymbol{\nabla} \u_1 \right) \right].
\end{align}
Fourier transforming Eq.~(\ref{eq:constitutive2}) and using Eq.~(\ref{eq:constitutiveF1}), we obtain the two harmonics as
\begin{align}\label{eq:2ndorder2}
&\left(1+2i \De_1\right) \tilde\btau^{(2)}_2-\left( 1+2i\De_2\right) \tilde\gamd^{(2)}_2 \notag \notag  \\
&= \frac{1}{2} \frac{\De_2-\De_1}{1+i \De_1}\left[\tilde\u_1 \cdot \boldsymbol{\nabla} \tilde\gamd_1- \left(^{t}\boldsymbol{\nabla} \tilde\u_1 \cdot \tilde\gamd_1 +\tilde\gamd_1 \cdot \boldsymbol{\nabla} \tilde\u_1 \right) \right],
\end{align}
and
\begin{align}\label{eq:2ndorder0}
&\tilde\btau^{(0)}_2-\tilde\gamd^{(0)}_2 \notag \\
&= \frac{1}{2} \frac{\De_2-\De_1}{1+i \De_1} \left[\tilde\u_1^{*} \cdot \boldsymbol{\nabla} \tilde\gamd_1- \left(^{t}\boldsymbol{\nabla} \tilde\u_1^{*}  \cdot  \tilde\gamd_1 +\tilde\gamd_1 \cdot \boldsymbol{\nabla} \tilde\u_1^{*} \right) \right],
\end{align}
where the starred variables denote complex conjugates in this paper.
Finally, taking the divergence and then curl of both Eq.~(\ref{eq:2ndorder2}) and Eq.~(\ref{eq:2ndorder0}), and using the knowledge of the first-order solution, we obtain the equation for the second-order streamfunctions as simply
\begin{subequations}
\begin{align}
\boldsymbol{\nabla}^4 \tilde\Psi_2^{(2)} =0, \\
\boldsymbol{\nabla}^4 \tilde\Psi_2^{(0)} =0.
\end{align}
\end{subequations}

\subsubsection{Boundary conditions}
The boundary condition at this order is given by
\begin{align}
\u_2 = - \Theta \frac{\p \u_1}{\p \theta} -r \Theta \dot{\Theta} \langle \bf{t} \rangle,
\end{align}
when evaluated at $\theta = \pi/2$.
In Fourier notation and with the first-order solution, the boundary conditions for the second-order average flow and the second harmonic read
\begin{subequations}
\begin{align}
\tilde\u_2^{(0)} &=0, \\
\tilde\u_2^{(2)} &= -\frac{ir}{2} \langle t \rangle.
\end{align}
\end{subequations}
In addition, the no-slip and no-penetration boundary condition are imposed at $\theta=0$.

\subsubsection{Solution}
The solution satisfying the second-order equation and the boundary conditions is given by 
\begin{subequations}
\begin{align}
\tilde\Psi_2^{(0)}&=0,\\
\tilde\Psi_2^{(2)} &= \frac{ir}{4} \left(\frac{1}{2} \sin 2\theta-\frac{\pi}{4} \cos2\theta-\theta+\frac{\pi}{4} \right),\\
\tilde{u}^{(2)}_{2r} &= \frac{ir}{4} \left( \cos2\theta+\frac{\pi}{2} \sin2\theta -1\right),\\
\tilde{u}^{(2)}_{2\theta} &= -\frac{ir}{4} \left( \sin2\theta-\frac{\pi}{2} \cos2\theta-2\theta+\frac{\pi}{2}\right).
\end{align}
\end{subequations}
As anticipated, there is no time-averaged flow at second order, and we  proceed with calculations at higher order.

\subsection{Third-order solution}
\subsubsection{Governing equation}
The third-order Oldroyd-B relation is given by 
\begin{align}\label{eq:constitutive3}
&\left( 1+\De_1 \frac{\p }{\p t}\right) \btau_3 -\left( 1+\De_2 \frac{\p }{\p t}\right) \gamd_3 \notag\\ 
&=  \De_2 \left[\u_1 \cdot \boldsymbol{\nabla} \gamd_2- \left(^{t} \boldsymbol{\nabla} \u_1 \cdot \gamd_2+\gamd_2 \cdot \boldsymbol{\nabla} \u_1 \right) \right] \notag \\
& - \De_1  \left[\u_2 \cdot \boldsymbol{\nabla} \btau_1- \left(^{t} \boldsymbol{\nabla} \u_2 \cdot \btau_1+\btau_1 \cdot \boldsymbol{\nabla} \u_2 \right) \right] \notag \\
& +  \De_2 \left[\u_2 \cdot \boldsymbol{\nabla} \gamd_1- \left(^{t} \boldsymbol{\nabla} \u_2 \cdot \gamd_1+\gamd_1 \cdot \boldsymbol{\nabla} \u_2 \right) \right] \notag \\
& - \De_1  \left[\u_1 \cdot \boldsymbol{\nabla} \btau_2- \left(^{t} \boldsymbol{\nabla} \u_1 \cdot \btau_2+\btau_2 \cdot \boldsymbol{\nabla} \u_1 \right) \right] .
\end{align}
Aiming at obtaining the average flow at $O(\epsilon^4)$, we only need to calculate the first harmonic at  $O(\epsilon^3)$ since the third harmonic will only enter the oscillatory part at $O(\epsilon^4)$ (see the fourth-order calculations for details).
Therefore, upon Fourier transform, we obtain the first harmonic component of Eq.~(\ref{eq:constitutive3}) as
\begin{align} \label{eq:constitutiveF3}
&\left(1+i \De_1\right) \tilde\btau^{(1)}_3-\left( 1+i\De_2\right) \tilde\gamd^{(1)}_3 \notag \\
&= \frac{1}{2} \frac{\De_2-\De_1}{1+2i \De_1}\left[\tilde\u_1^{*} \cdot \boldsymbol{\nabla} \tilde\gamd_2^{(2)}- \left(^{t}\boldsymbol{\nabla} \tilde\u_1^{*} \cdot \tilde\gamd_2^{(2)} +\tilde\gamd_2^{(2)} \cdot \boldsymbol{\nabla} \tilde\u_1^{*} \right) \right] \notag \\
&+ \frac{1}{2} \frac{\De_2-\De_1}{1-i \De_1}\left[\tilde\u_2^{(2)} \cdot \boldsymbol{\nabla} \tilde\gamd_1^{*}- \left(^{t}\boldsymbol{\nabla} \tilde\u_2^{(2)} \cdot \tilde\gamd_1^{*} +\tilde\gamd_1^{*} \cdot \boldsymbol{\nabla} \tilde\u_2^{(2)} \right) \right] ,
\end{align}
where we have used the constitutive relations given by Eqs.~(\ref{eq:constitutiveF1}) and (\ref{eq:2ndorder2}). 
Taking the divergence and then curl of Eq.~(\ref{eq:constitutiveF3}), we obtain the equation for the first harmonic of the third-order streamfunction
\begin{align}
\boldsymbol{\nabla}^4 \tilde\Psi^{(1)}_3 = \frac{3i \De_1 (\De_1-\De_2)\,  \cos2\theta}{r^2(1-i\De_1)(1+2i\De_1)(1+i\De_2)}\cdot
\end{align}

\subsubsection{Boundary condition}
At $\theta = \pi/2$, the boundary condition at this order is given by 
\begin{align}
\u_3 = -\Theta \frac{\p \u_2}{\p \theta} - \frac{1}{2} \Theta^2 \frac{\p^2 \u_1}{\p \theta^2}-\frac{1}{2} r \dot{\Theta} \Theta^2 \langle \n \rangle,
\end{align}
evaluating at $\theta = \pi/2$. The boundary condition for the first harmonic component, in Fourier space, is then given by 
\begin{align}
\tilde\u_3^{(1)}=\frac{ir\pi}{8} \langle \t \rangle -\frac{ir}{4} \langle \n \rangle.
\end{align}
At $\theta=0$, we also have the no-slip and no-penetration boundary condition.

\subsubsection{Solution}
The solution at this order has the form
\begin{widetext}
\begin{subequations}
\begin{align}
\tilde\Psi_3^{(1)} = \ &\frac{r^{2}}{2} \Biggl[ \left(\frac{\pi^{2}}{8}\alpha+\frac{\pi^{2}-4}{32}i\right)\cos2\theta
-\frac{\pi}{4}\left(\alpha+\frac{i}{4}\right)\sin2\theta
+\frac{\pi}{2}\left(\alpha+\frac{i}{4}\right)\theta 
+\left(\frac{4-\pi^{2}}{32}i -\frac{\pi^{2}}{8}\alpha\right)   \notag \\ 
&+\alpha\theta\sin2\theta
\Biggr],\\
\tilde u_{3,r}^{(1)} = \ &\frac{r}{2} \Biggl[ \left(\frac{4-\pi^{2}}{16}i -\frac{\pi^{2}}{4}\alpha\right)\sin2\theta-\frac{\pi}{2}\left(\alpha+\frac{i}{4}\right)\cos2\theta+\frac{\pi}{2}\left(\alpha+\frac{i}{4}\right)
+\alpha(2\theta \cos2\theta+\sin2\theta)\Biggr], \\
\tilde u_{3,\theta}^{(1)}= \ &-r \Biggl[ \left(\frac{\pi^{2}}{8}\alpha+\frac{\pi^{2}-4}{32}i\right)\cos2\theta
-\frac{\pi}{4}\left(\alpha+\frac{i}{4}\right)\sin2\theta
+\frac{\pi}{2}\left(\alpha+\frac{i}{4}\right)\theta
+\left(\frac{4-\pi^{2}}{32}i -\frac{\pi^{2}}{8}\alpha\right)  \notag \\
&+\alpha\theta\sin2\theta
\Biggr],
\end{align}
\end{subequations}
\end{widetext}
where we have defined the constant
\begin{align}
\alpha=\frac{3i\De_1(\De_2-\De_1)}{8(1-i\De_1)(1+2i\De_1)(1+i\De_2)}\cdot
\end{align}

\subsection{Fourth-order solution}
\subsubsection{Governing equation}
Finally, the fourth-order Oldroyd-B relation is given by
\begin{align}\label{eq:constitutive4}
&\left( 1+\De_1 \frac{\p }{\p t}\right) \btau_4 -\left( 1+\De_2 \frac{\p }{\p t}\right) \gamd_4 \notag\\ 
&=  \De_2 \left[\u_1 \cdot \boldsymbol{\nabla} \gamd_3- \left(^{t} \boldsymbol{\nabla} \u_1 \cdot \gamd_3+\gamd_3 \cdot \boldsymbol{\nabla} \u_1 \right) \right] \notag \\
& - \De_1  \left[\u_1 \cdot \boldsymbol{\nabla} \btau_3- \left(^{t} \boldsymbol{\nabla} \u_1 \cdot \btau_3+\btau_3 \cdot \boldsymbol{\nabla} \u_1 \right) \right] \notag \\
& +  \De_2 \left[\u_2 \cdot \boldsymbol{\nabla} \gamd_2- \left(^{t} \boldsymbol{\nabla} \u_2 \cdot \gamd_2+\gamd_2 \cdot \boldsymbol{\nabla} \u_2 \right) \right] \notag \\
& - \De_1  \left[\u_2 \cdot \boldsymbol{\nabla} \btau_2- \left(^{t} \boldsymbol{\nabla} \u_2 \cdot \btau_2+\btau_2 \cdot \boldsymbol{\nabla} \u_2 \right) \right] \notag \\
& +  \De_2 \left[\u_3 \cdot \boldsymbol{\nabla} \gamd_1- \left(^{t} \boldsymbol{\nabla} \u_3 \cdot \gamd_1+\gamd_1 \cdot \boldsymbol{\nabla} \u_3 \right) \right] \notag \\
& - \De_1  \left[\u_3 \cdot \boldsymbol{\nabla} \btau_1- \left(^{t} \boldsymbol{\nabla} \u_3 \cdot \btau_1+\btau_1 \cdot \boldsymbol{\nabla} \u_3 \right) \right] .
\end{align}
Since we wish to characterize  the time-average flow, $\tilde\u_4^{(0)}$, we calculate the time-average of Eq.~\eqref{eq:constitutive4} and obtain
\begin{align} \label{eq:constitutiveF4}
&\tilde{\bold{\btau}}^{(0)}_4-\tilde{{\gamd}}^{(0)}_4 \notag \\ 
&= \frac{1}{2} \De_2 \left[  \tilde{\bold{u}}^{*}_1\cdot \boldsymbol{\nabla} \tilde{\gamd}^{(1)}_3 - \left( {}^t \boldsymbol{\nabla} \tilde{\bold{u}}^{*}_1 \cdot \tilde{\gamd}^{(1)}_3+\tilde{\gamd}^{(1)}_3 \cdot \boldsymbol{\nabla} \tilde{\bold{u}}^{*}_1 \right)   \right] \notag\\ 
&-\frac{1}{2} \De_1 \left[  \tilde{\bold{u}}^{*}_1\cdot \boldsymbol{\nabla} \tilde{\btau}^{(1)}_3 - \left( {}^t \boldsymbol{\nabla} \tilde{\bold{u}}^{*}_1 \cdot \tilde{\btau}^{(1)}_3+\tilde{\btau}^{(1)}_3 \cdot \boldsymbol{\nabla} \tilde{\bold{u}}^{*}_1 \right)   \right] \notag\\   
&+\frac{1}{2} \De_2 \left[  \tilde{\bold{u}}^{(2)*}_2\cdot \boldsymbol{\nabla} \tilde{\gamd}^{(2)}_2 - \left( {}^t \boldsymbol{\nabla} \tilde{\bold{u}}^{(2)*}_2 \cdot \tilde{\gamd}^{(2)}_2+\tilde{\gamd}^{(2)}_2 \cdot \boldsymbol{\nabla} \tilde{\bold{u}}^{(2)*}_2 \right)   \right] \notag\\  
&-\frac{1}{2} \De_1 \left[  \tilde{\bold{u}}^{(2)*}_2\cdot \boldsymbol{\nabla} \tilde{\btau}^{(2)}_2 - \left( {}^t \boldsymbol{\nabla} \tilde{\bold{u}}^{(2)*}_2 \cdot \tilde{\btau}^{(2)}_2+\tilde{\btau}^{(2)}_2 \cdot \boldsymbol{\nabla} \tilde{\bold{u}}^{(2)*}_2 \right)   \right] \notag\\  
&+\frac{1}{2} \De_2 \left[  \tilde{\bold{u}}^{(1)*}_3\cdot \boldsymbol{\nabla} \tilde{\gamd}_1 - \left( {}^t \boldsymbol{\nabla} \tilde{\bold{u}}^{(1)*}_3 \cdot \tilde{\gamd}_1+\tilde{\gamd}_1 \cdot \boldsymbol{\nabla} \tilde{\bold{u}}^{(1)*}_3 \right)   \right] \notag\\  
&-\frac{1}{2} \De_1 \left[  \tilde{\bold{u}}^{(1)*}_3\cdot \boldsymbol{\nabla} \tilde{\btau}_1 - \left( {}^t \boldsymbol{\nabla} \tilde{\bold{u}}^{(1)*}_3 \cdot \tilde{\btau}_1+\tilde{\btau}_1 \cdot \boldsymbol{\nabla} \tilde{\bold{u}}^{(1)*}_3 \right)   \right].
\end{align}
As done previously, we take the divergence and then curl of Eq.~(\ref{eq:constitutiveF4}), and invoke the lower-order constitutive relations Eqs.~(\ref{eq:constitutiveF1}),~(\ref{eq:2ndorder2}) and (\ref{eq:constitutiveF3}), to obtain the equation for streamfunction of the average flow
\begin{align}
\boldsymbol{\nabla}^4 \tilde\Psi^{(0)}_4 = \beta \frac{A_4 \sin 2\theta+ B_4 \cos 2\theta + C_4 \sin 4\theta }{r^2},
\end{align} 
where 
\begin{align}
\beta &= \frac{\De_2-\De_1}{2(1+De^2_1)(2\De_1-i)},\\
A_4 &=8 \alpha-\De_1+8 i \alpha \De_1+4 i \De_1^2+16 \alpha \De_1^2,\\
B_4 &= 2 \pi \left(1+i\De_1+2\De_1^2)(\alpha+\alpha^{*}\right),\\
C_4 &= -2 \bigl( 8\alpha+8i\alpha\De_1+3i\De_1^2+16\alpha\De_1^2 \notag \\
& \ \ \ +4\alpha^{*}+4i\De_1\alpha^{*}+8\De_1^2 \alpha^{*} \bigr).
\end{align}

\subsubsection{Boundary conditions}
At $\theta = \pi/2$, the boundary condition at this order is written as
\begin{align}
\u_4 = -\Theta \frac{\p \u_3}{\p \theta}-\frac{1}{2} \Theta^2 \frac{\p^2 \u_2}{\p \theta^2}-\frac{1}{6} \Theta^3 \frac{\p^3 \u_1}{\p \theta^3}+\frac{1}{6} r \dot{\Theta} \Theta^3 \langle \t \rangle,
\end{align}
which we then  Fourier-transform to obtain the boundary condition for $\tilde\u^{(0)}_4$. In addition, since we are only interested in the time-averaged flow, \textit{i.e.}, real part of the solution $\Real\{\tilde\u^{(0)}_4\}$, the boundary condition at $\theta=\pi/2$ can be simplified as  
\begin{align}
\Real\{\tilde\u^{(0)}_4\} = \frac{r\left(8-\pi^2\right)}{8} \Real\{\alpha\} \langle \t \rangle,
\end{align}
where 
\begin{align}
\Real\{\alpha\} = \frac{-3\De_1(\De_1-\De_2)(\De_1+\De_2+2\De^2_1\De_2)}{8(1+\De^2_1)(1+4\De^2_1)(1+\De^2_2)}\cdot
\end{align}
Finally, as usual, we have the no-slip and no-penetration boundary conditions at $\theta=0$.

\subsubsection{Solution}
Solving the inhomogeneous biharmonic equation with the boundary conditions above, we obtain our main result, namely the analytical formula for the time-averaged flow as 
\begin{widetext}
\begin{subequations}\label{eq:order4sol}
\begin{align}
\Real\left\{\tilde{\Psi}^{(0)}_{4}\right\} =&
\frac{r^2 \De_1 \left(\De_1-\De_2\right)\left(2 \De_2 \De_1^2+\De_1+\De_2\right)}{512 \left(\De_1^2+1\right){}^2 \left(4 \De_1^2+1\right) \left(\De_2^2+1\right)} \notag\\
&\Bigl[32 \pi -3 \pi ^3+8 \pi  \De_1 \De_2-3 \pi ^3 \De_1 \De_2+24 \pi  \De_1^2+4 \theta  \bigl(-20+3 \pi ^2 \notag \\
&-12 \De_1^2-8 \De_1 \De_2+3 \pi ^2 \De_1 \De_2 \bigr) +\sin 4\theta \left(-4-12 \De_1^2+8 \De_1 \De_2\right) \notag \\
&+\cos 2\theta \left(-32 \pi +3 \pi ^3+48 \theta -24 \pi  \De_1^2+48 \theta  \De_1^2-8 \pi  \De_1 \De_2+3 \pi ^3 \De_1 \De_2\right) \notag\\
&+\sin 2\theta \left(24-6 \pi ^2+24 \De_1^2-24 \pi  \theta  \De_1^2-6 \pi ^2 \De_1 \De_2+24 \pi  \theta  \De_1 \De_2\right) \Bigr],\\
\Real\left\{\tilde{u}^{(0)}_{4r}\right\} =&
\frac{r \De_1 \left(\De_2-\De_1\right) \left(2 \De_2 \De_1^2+\De_1+\De_2\right)}{256 \left(\De_1^2+1\right){}^2 \left(4 \De_1^2+1\right) \left(\De_2^2+1\right)} \notag\\
& \Bigl[ 40-6 \pi ^2+24 \De_1^2+16 \De_1 \De_2-6 \pi ^2 \De_1 \De_2+\cos 4\theta \left(8+24 \De_1^2-16 \De_1 \De_2\right)\notag\\
&+\sin 2\theta \left(-32 \pi +3 \pi ^3+48 \theta -12 \pi  \De_1^2+48 \theta  \De_1^2-20 \pi  \De_1 \De_2+3 \pi ^3 \De_1 \De_2\right)\notag\\
&+\cos 2\theta \left(-48+6 \pi ^2-48 \De_1^2+24 \pi  \theta  \De_1^2+6 \pi ^2 \De_1 \De_2-24 \pi  \theta  \De_1 \De_2\right) \Bigr],\\
\Real\left\{\tilde{u}^{(0)}_{4\theta}\right\} =&
\frac{r \De_1 \left(\De_2-\De_1\right) \left(2 \De_2 \De_1^2+\De_1+\De_2\right)}{256 \left(\De_1^2+1\right){}^2 \left(4 \De_1^2+1\right) \left(\De_2^2+1\right)} \notag\\
&\Bigl[ 32 \pi -3 \pi ^3+8 \pi  \De_1 \De_2-3 \pi ^3 \De_1 \De_2+24 \pi  \De_1^2+4 \theta  \bigl(-20+3 \pi ^2 \notag \\
&-12 \De_1^2-8 \De_1 \De_2+3 \pi ^2 \De_1 \De_2\bigr)+\sin 4\theta \left(-4-12 \De_1^2+8 \De_1 \De_2\right) \notag \\
&+\cos 2\theta \left(-32 \pi +3 \pi ^3+48 \theta -24 \pi  \De_1^2+48 \theta  \De_1^2-8 \pi  \De_1 \De_2+3 \pi ^3 \De_1 \De_2\right) \notag\\
&+\sin 2\theta \left(24-6 \pi ^2+24 \De_1^2-24 \pi  \theta  \De_1^2-6 \pi ^2 \De_1 \De_2+24 \pi  \theta  \De_1 \De_2\right)\Bigr]. 
\end{align}
\end{subequations}
\end{widetext}

%%%%%%%%%%%%%%%%%%%%%%%%%%%%%%%%%%%%%%%%
%%%%%%%%%%%%%%%%%%%%
%%%% RESULTS
%%%%%%%%%%%%%%%%%%%%
%%%%%%%%%%%%%%%%%%%%%%%%%%%%%%%%%%%%%%%%

\begin{figure}
\centering
\includegraphics[width=0.7\textwidth]{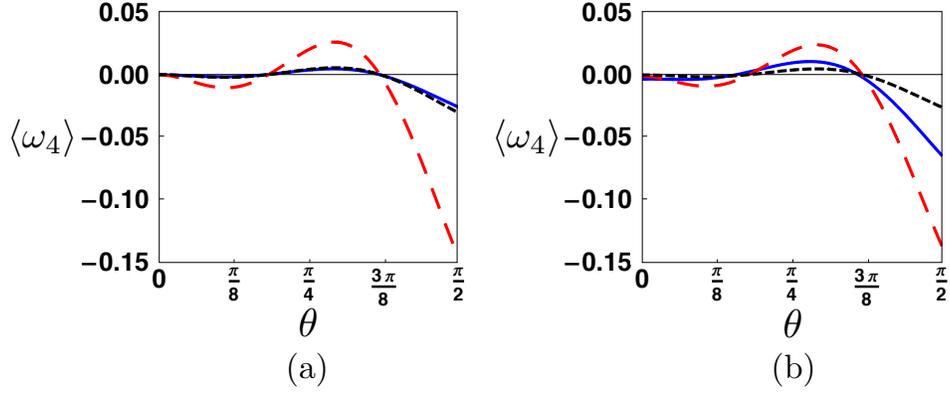}
\caption{\label{fig:vorticity}Time-averaged vorticity, $\langle \omega_4\rangle$, as a function of polar  angle $\theta$. 
(a): Fixed Deborah number ($\De =100$) and  $\eta_s / \eta =$ 0.1 (blue  solid line), 0.01 (red  dashed lines) and 0.001 (black  dotted line); 
(b):  Fixed relative viscosity ($\eta_s/\eta = 0.1$) and $\De$ = 1 (blue solid line), 10 (red  dashed line), and 100 (black  dotted line).}
\end{figure}

\section{\label{sec:results}Characterization of the time-averaged flow}

In the  analysis above, we have computed the flow field perturbatively up to order $O(\epsilon^4)$ and found that a nonzero time-averaged flow occurs at that order, as described by  Eq.~(\ref{eq:order4sol}). Hereafter, for convenience, we rewrite the two Deborah numbers as $\De_1=\De$ and $\De_2=\De \zeta$, where $\zeta = \eta_s /\eta$ is the relative viscosity of the solvent vs. total fluid. 
The creation of a net flow by the  tethered flapping motion demonstrates explicitly that Purcell's scallop theorem breaks down in a viscoelastic fluid. This suggests that reciprocal flapping-like motion can be exploited for pumping polymeric fluids in simple geometries even in the absence of inertia -- a situation which is impossible  in Newtonian fluids.  In the following sections, we explore the properties of this time-average flow and its dependance on both the actuation frequency and material properties of the fluid.

\subsection{Streamline and vorticity pattern}
\begin{figure*}[t]
\centering
\includegraphics[width=0.8\textwidth]{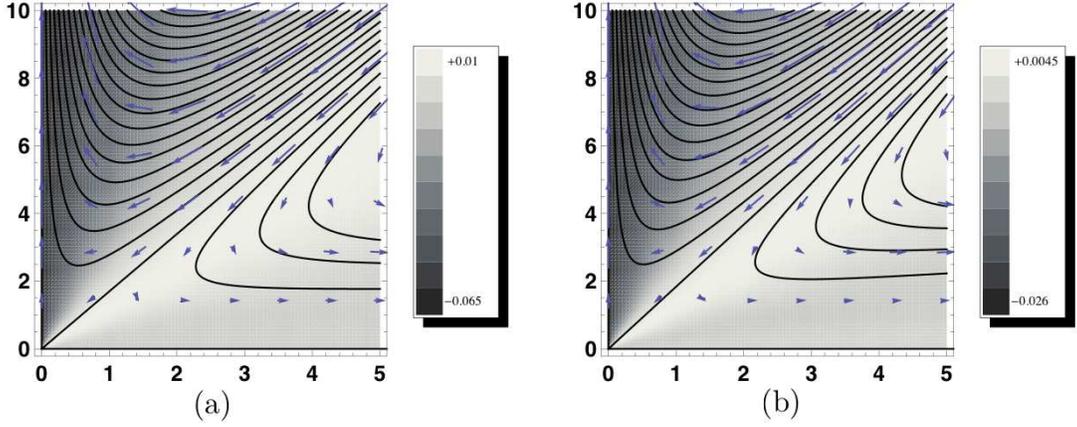}
\caption{\label{fig:streamline} Streamline and vorticity pattern for $\eta_s / \eta = 0.1$; (a):   
 $\De=1$; (b):  $\De =100$. The grayscale map displays the value of the vorticity, with legend shown on the right of each plot.}
\end{figure*}

With the streamfunction explicitly calculated, we can easily compute the  flow streamlines, as well as the flow  vorticity, given by $\langle \omega_4 \rangle = - \boldsymbol{\nabla}^2  \langle \Psi_4 \rangle$, or
\begin{align}
\langle \omega_4 \rangle =& \frac{\De^3 (1-\zeta ) \left(1+\zeta +2 \De^2 \zeta \right)}{128 \left(1+\De^2\right)^2 \left(1+4 \De^2\right) \left(1+\De^2 \zeta ^2\right)} \notag \\
&\Bigl[ -32 \pi -24 \De^2 \pi +3 \pi ^3-8 \De^2 \pi  \zeta +3 \De^2 \pi ^3 \zeta \notag\\
&-4 \left(-20-12 \De^2+3 \pi ^2-8 \De^2 \zeta +3 \De^2 \pi ^2 \zeta \right) \theta  \notag \\
& +\left(24 \De^2 \pi -24 \De^2 \pi  \zeta \right) \cos2 \theta +\left(48+48 \De^2\right) \sin2 \theta \notag\\
&+\left(-12-36 \De^2+24 \De^2 \zeta \right) \sin4\theta \Bigr].
\end{align}
We see that the vorticity is only a function of the polar angle $\theta$ between the wall and the flapper. The vorticity is plotted as a function of the angle $\theta$ in Fig.~\ref{fig:vorticity} for different relative viscosities (Fig.~\ref{fig:vorticity}a) and different Deborah numbers (Fig.~\ref{fig:vorticity}b). The locations where the vorticity changes its sign are apparently  invariant and occur around $\theta \approx 3\pi/16$ (from negative to positive) and $\theta \approx 3\pi/8$ (from positive to negative). The streamline pattern and vorticity distribution are also qualitatively similar for different Deborah numbers and relative viscosities, as illustrated in  Fig.~\ref{fig:streamline} for different Deborah numbers ($\De=1$ and $\De=100$) at a fixed relative viscosity of $0.1$. It can be noted that, keeping the relative viscosity fixed, increasing the Deborah number leads to more inclined streamlines (greater vertical velocity components) near the flat wall ($\theta=0$).

\begin{figure}
\centering
\includegraphics[width=0.7\textwidth]{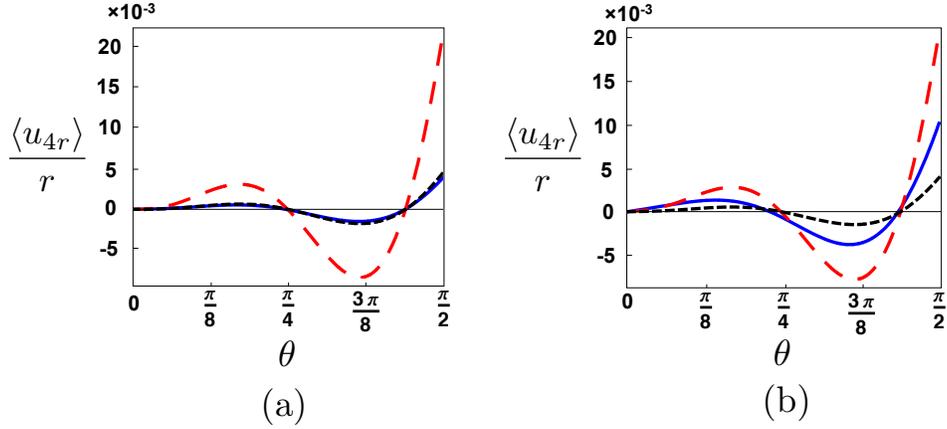}
\caption{\label{fig:direction}Net flow velocity along the average flapper position, $\langle u_{4r}\rangle/r$, as a function of the polar angle $\theta$. 
(a):  fixed Deborah number ($\De=100$) and $\eta_s / \eta =$ 0.1 (blue solid line), 0.01 (red  dashed line) and 0.001 (black dotted line);
(b) fixed relative viscosity ($\eta_s/\eta =$ 0.1) and De = 1 (blue solid line), 10 (red dashed line), and 100 (black dotted line).}
\end{figure}

\subsection{Directionality of the flow}
As shown from the arrows in the streamline pattern in Fig.~\ref{fig:streamline}, the flapping motion draws the polymeric fluid towards the hinge point at an acute angle, and pumps the fluid away from the hinge point along both the flat wall and the average flapper position. To illustrate this directionality further, we plot in Fig.~\ref{fig:direction} the radial velocity per unit radius against the polar angle  for different relative viscosities (Fig.~\ref{fig:direction}a) and Deborah numbers (Fig.~\ref{fig:direction}b). Again, the locations where the radial velocity changes its sign are apparently invariant under the change of relative viscosity or Deborah number, and occur around $\theta \approx \pi/4$ (positive to negative) and $\theta \approx 7\pi/16$ (negative to positive).

\section{\label{sec:optimization}Optimization}
Having identified the basic flow patterns generated by the flapping motion, we now turn to a possible  optimization of the pumping performance. Specifically, we address the question: what is the optimal Deborah number at which the largest flow can be generated? Since different optimality criteria can be defined, we consider here three  different ``optimality measures'' for  the net flow, and show they all generate essentially the same conclusion.

\subsubsection{Flow along the boundary}
\begin{figure*}
\centering
\includegraphics[width=0.7\textwidth]{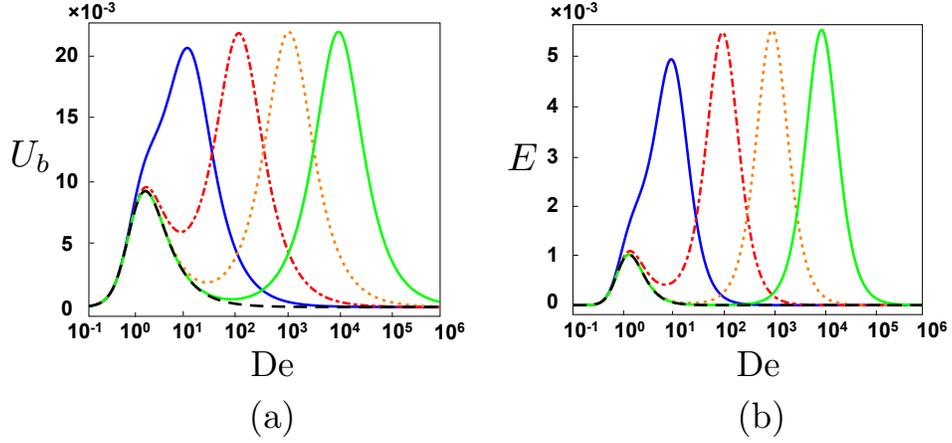}
\caption{\label{fig:optim}Dependence of pumping performance with the Deborah number, for two different pumping measures. 
(a):  Reduced flow velocity along the average flapper position, $U_b$; 
(b): Reduced kinetic energy, $E$. 
For both cases:  $\eta_s/\eta=0.1$ (left solid line, blue); $\eta_s/\eta=0.01$ (red dot-dashed line); $\eta_s/\eta=0.001$ (orange dotted line); $\eta_s/\eta=0.0001$ (right solid line, green); $\eta_s/\eta=0$ (black dashed line).}
\end{figure*}

Since the flapping motion pumps the fluid away from the hinge point along the average flapper position, one natural measure of the pumping performance is the magnitude of flow along the average flapper position ($\theta=\pi/2$). Note that the velocity field is directly proportional to the radius, and recall that the velocity is only radial along the average flapper position as required by symmetry. Consequently, the dependence of the intrinsic flow strength upon the Deborah number can be characterized by the ratio between the radial velocity along the average flapper position  and the radial distance, 
\begin{align}\label{eq:optimB}
U_b \left(\De, \zeta \right)=& \ \frac{\langle u_{4r} \rangle (r, \theta=\pi/2)}{r} \notag\\
=& \ \frac{3 \De^3 \left(\pi ^2-8\right) (1-\zeta ) \left(1+\zeta +2 \De^2 \zeta \right)}{64 \left(1+5 \De^2+4 \De^4\right) \left(1+\De^2 \zeta ^2\right)},
\end{align}
which is plotted for different relative viscosities in Fig.~\ref{fig:optim}a.  {From Eq.~(\ref{eq:optimB}), we see that  for small values of $\De$, $U_b \sim \De^3$,  whereas for large values of $\De$, $U_b\sim1/\De$, and therefore an optimal Deborah number is expected to exist.} This is confirmed in  Fig.~\ref{fig:optim}a, where we see that for each value of the relative viscosity, there is an optimal value of the Deborah number where the flow along the boundary is maximal. 
For small relative viscosities, we note the presence of  two local peaks (in contrast,  only one exists for $\eta_s/ \eta =  10^{-1}$). Physically, by decreasing the relative viscosity, we are varying the retardation time of the fluid, while keeping the relaxation time fixed. The position of the second peak changes correspondingly and commensurately when the relative viscosity is varied by orders of magnitude, while the position the first peak is unchanged. When the relative viscosity is set to zero (zero retardation time, which is a singular limit), we see in Fig.~\ref{fig:optim}a a  single peak at  essentially the same Deborah number as before. From these observations, we deduce that the two local optimal Deborah numbers arise  from two different properties of the fluid, respectively relaxation and retardation.  For small relative viscosities, the small local optimal Deborah number can be attributed to relaxation while the larger local optimal Deborah number can be attributed to  retardation, and it disappears in the singular (and unphysical) limit of zero retardation.

\subsubsection{Kinetic energy}
Another possible  optimization measure is related to the total kinetic energy of the average flow. Since the velocity field is directly proportional to the radius, it takes the general form $\langle u_{4r} \rangle = r f(\theta)$ and  $\langle u_{4\theta} \rangle = r g(\theta)$, where the functions $f(\theta)$ and $g(\theta)$ can be found from Eq.~(\ref{eq:order4sol}). Therefore, the dependence of the total kinetic energy of the average flow upon the Deborah number can be characterized by a reduced energy given by the integral over the polar angle
\begin{align}
E(\De,\zeta) =\int^{\pi/2}_0 \left[f(\theta) \right]^2 +\left[g(\theta) \right]^2d\theta,
\end{align}
and is given analytically by 
\begin{align}\label{eq:optimE}
E(\De,\zeta)=& \frac{\De^6 \pi  (-1+\zeta )^2 \left(1+\zeta +2 \De^2 \zeta \right)^2 }{98304 \left(1+\De^2\right)^4 \left(1+4 \De^2\right)^2 \left(1+\De^2 \zeta ^2\right)^2} \notag \\
&\Bigl[-2074-4028 \De^2-2058 \De^4+1054 \pi ^2 \notag\\
&+858 \De^2 \pi ^2-144 \De^4 \pi ^2-174 \pi ^4 -45 \De^2 \pi ^4 \notag\\
&+36 \De^4 \pi ^4+9 \pi ^6-120 \De^2 \zeta +88 \De^4 \zeta  \notag\\
&+1250 \De^2 \pi ^2 \zeta +1146 \De^4 \pi ^2 \zeta -303 \De^2 \pi ^4 \zeta \notag\\
&-117 \De^4 \pi ^4 \zeta +18 \De^2 \pi ^6 \zeta -104 \De^4 \zeta ^2 \notag\\
&+52 \De^4 \pi ^2 \zeta ^2-93 \De^4 \pi ^4 \zeta ^2 +9 \De^4 \pi ^6 \zeta ^2 \Bigr].
\end{align}
{With a fixed relative viscosity, for small values of $\De$, we have $E \sim \De^6$ , whereas $E \sim 1/\De^2$ for large values of $\De$, so an optimal $\De$ should exist.} The function $E(\De,\zeta)$ is plotted for different values of the relative viscosity in Fig.~\ref{fig:optim}b, and similarly to the previous section we see indeed the existence of an optimal value of $\De$ for each $\zeta$.

\subsubsection{Enstrophy}
Finally, we also consider the dependence of the enstrophy of the flow upon the Deborah number. The total enstrophy of the flow is proportional to the integral
\begin{align}
\mathcal{E}(\De, \zeta) = \int^{\pi/2}_{0} \omega^2_{4}d\theta,
\end{align}
which can be analytically calculated  to be
\begin{align}
\mathcal{E} =& \frac{\De^6 \pi  (-1+\zeta )^2 \left(1+\zeta +2 \De^2 \zeta \right)^2 }{98304 \left(1+\De^2\right)^4 \left(1+4 \De^2\right)^2 \left(1+\De^2 \zeta ^2\right)^2} \notag \\
& \Bigl[-1800-14256 \De^2-12744 \De^4+616 \pi ^2 \notag \\
&+2424 \De^2 \pi ^2 +1440 \De^4 \pi ^2-120 \pi ^4-72 \De^2 \pi ^4 \notag \\
&+9 \pi ^6 +10656 \De^2 \zeta +11232 \De^4 \zeta -1192 \De^2 \pi ^2 \zeta \notag \\
&-456 \De^4 \pi ^2 \zeta -168 \De^2 \pi ^4 \zeta -72 \De^4 \pi ^4 \zeta \notag \\
&+18 \De^2 \pi ^6 \zeta -288 \De^4 \zeta ^2-368 \De^4 \pi ^2 \zeta ^2 & \notag \\
&-48 \De^4 \pi ^4 \zeta ^2 +9 \De^4 \pi ^6 \zeta ^2\Bigr].
\end{align}
The variation  of $\mathcal{E}$ with $\De$ turns out to be very similar to the one
for $E$ (Eq.~\ref{eq:optimE}, shown  in Fig.~\ref{fig:optim}b), and is not reproduced here.

\subsubsection{Optimal Deborah number}
\begin{figure}
\centering
\includegraphics[width=0.4\textwidth]{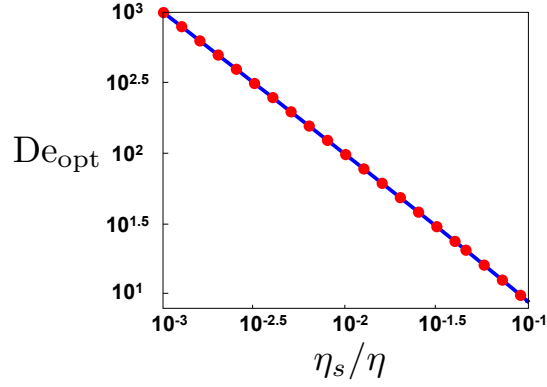}
\caption{\label{fig:loglogoptim}Optimal Deborah number, $\De_{\text{opt}}$, as a function of relative viscosity, $\eta_s/\eta$, by optimizing (a) the flow along the average position of the flapper $\theta = \pi/2$ (blue solid line) and (b) the flow kinetic energy (red dots).}
\end{figure}

We next compute numerically the optimal Deborah number, maximizing the pumping measures in both Eqs.~(\ref{eq:optimB}) and (\ref{eq:optimE}), as a function of the relative viscosity. The results are displayed in Fig.~\ref{fig:loglogoptim}. The optimality conditions turn out to quantitatively agree for both  pumping measures, and correspond to an inverse linear  relationship between $\De_{\text{opt}}$ and $\zeta$. This scaling is confirmed by an asymptotic analysis of the exact analytical formula for $\De_{\text{opt}}$ found by setting the partial derivative of Eq.~(\ref{eq:optimB}) to zero, and  showing that indeed $\De_{\text{opt}} \sim 1/\zeta$  for small values of $\zeta$.

%%%%%%%%%%%%%%%%%%
%%%%%%%%%%%%%%%%%%
% 		DISCUSSION
%%%%%%%%%%%%%%%%%%
%%%%%%%%%%%%%%%%%%

\section{\label{sec:discussion}Discussion}

% SUMMARY OF WORK
In this paper, we have considered what is arguably the simplest geometrical setup to demonstrate that net fluid pumping can be obtained from the purely sinusoidal forcing of a viscoelastic fluid. The fluid was modeled as an Oldroyd-B fluid both for simplicity and because of the physical relevance of the model.
The main result we obtained is the time-averaged flow, described by Eq.~(\ref{eq:order4sol}), generated by the reciprocally flapping motion. In accordance with the scallop theorem, setting the Deborah number to zero in Eq.~(\ref{eq:order4sol}) leads to no flow, but a net flow occurs for all nonzero values of $\De$. Our calculations allow us to demonstrate explicitly the breaking of the scallop theorem in the context of fluid pumping, and suggest the possibility of  taking advantage of the intrinsic  nonlinearities of complex fluids for  their transport. Physically, such flow is being driven by normal-stress differences arising in the fluid and due to the stretching of the polymeric microstructures by the background flow.
The calculation was done asymptotically for small-amplitude flapping, and the net flow occurs at fourth order. As in the classic work by Moffatt \cite{moffatt}, our results  should be understood as similarity solutions which are valid close enough to the fixed hinge point such that the inertial effects are negligible. The advantage of such theoretical treatment is that it allows us to obtain the entire flow field analytically, in particular the spatial structure of the flow,  and the dependance of the net pumping on  the actuation parameters (the flapping frequency) and  the material properties of the fluid (relaxation time and viscosities).  Taking advantage of these analytical results, we have been able to analytically optimize the pumping performance, and derive the optimal Deborah number as a function of the fluid ratio of solvent to total viscosity.
Although we have considered here the simplest geometrical and dynamical setup possible, the results motivate future work which will focus on the  flapping of three-dimensional finite-size appendages in polymeric fluids.

%BIO-RELEVANCE

We now turn to the relevance of our results to biological transport. 
In Newtonian fluids, only the non-reciprocal component of the motion of cilia -- {\it i.e.} the difference between their effective and recovery strokes -- affects fluid transport \cite{brennen, blake}. 
In contrast,  we show in this paper that the back-and-forth components of cilia motion, which is reciprocal, does influence transport in the case of  viscoelastic biological fluids. The effect is expected to be crucial since the typical  Deborah number in ciliary transport is large, and  elastic effects of the fluid are  therefore likely to be significant. For example, from rheological measurements \cite{hwang, gilboa, lai}, we know the relaxation time of respiratory mucus ranges between $\lambda \approx 30 - 100$ s, and that of the cervical mucus present in female reproductive tract ranges from $\lambda \approx 1-100$ s  \cite{six,hwang}. In addition,  cilia typically oscillate at frequencies of $f=\omega/2\pi \approx 5 - 50$ Hz \cite{brennen}, and therefore,  ciliary transport of mucus  occurs at large (or very large) Deborah numbers, $\De=\lambda \omega \sim 10$ to $10^4$. 

In addition, the results of our paper should be contrasted with previous work. It was shown in Ref.~\cite{lauga1} that the presence of polymeric stresses leads to a decrease of the speed at which a fluid is pumped by a waving sheet  -- in that case a complex fluid led therefore to a degradation of the transport performance. In contrast, we demonstrate in the current paper a mode of actuation which is rendered effective by the presence of polymeric stresses -- the complex fluid leads therefore in this case to an improvement of the transport performance. For a general actuation gait, it is therefore not known a priori whether the presence of a complex fluid will lead to a degradation or an improvement of the pumping performance, and whether or not a general classification depending on the type of actuation gait can be derived remains a question to be addressed in the future.

\section*{Acknowledgments}
Funding by the National Science Foundation (grant CBET-0746285 to EL) is  gratefully acknowledged. 

\bibliography{Flapping}
\end{document}